\documentclass[prl,twocolumn,superscriptaddress,longbibliography,nofootinbib]{revtex4-1}

\usepackage[english]{babel}
\usepackage[utf8x]{inputenc}
\usepackage[T1]{fontenc}
\usepackage{csquotes}
\usepackage{url} 
\usepackage[a4paper,top=3cm,bottom=2cm,left=3cm,right=3cm,marginparwidth=1.75cm]{geometry}

\usepackage{amsmath}
\usepackage{amsfonts}
\usepackage{graphicx}
\usepackage[colorinlistoftodos]{todonotes}
\usepackage[colorlinks=true, allcolors=blue]{hyperref}

\usepackage[switch,columnwise]{lineno}
\usepackage{xcolor}

\def \quine #1{
\mbox{$\ulcorner$}
#1
\mbox{$\urcorner$}
}

\begin{document}

\title{Algorithmic measurement procedures}
\author{Aldo F. G. Solis-Labastida}
\affiliation{Instituto de Ciencias Nucleares, Universidad Nacional Aut\'onoma de M\'exico, Apdo. Postal 70-543, C.P. 04510  Cd. Mx., M\'exico}
\author{Jorge G. Hirsch}
\affiliation{Instituto de Ciencias Nucleares, Universidad Nacional Aut\'onoma de M\'exico, Apdo. Postal 70-543, C.P. 04510  Cd. Mx., M\'exico}

\begin{abstract}
 Measurements are shown to be processes designed to return figures: they are {\em effective}. 
This effectivity allows for a formalization as Turing machines, which can be described employing computation theory.  
Inspired in the {\em halting problem} we draw some limitations for measurement procedures: procedures that verify if a quantity is measured cannot work in every case. 

\end{abstract}
\maketitle

\section{Introduction}
Measurements play a central role in modern science.  
''In science ... facts are determined by observation or measurement of natural or experimental phenomena. 
A hypothesis is a proposed explanation of those facts. 
A theory is a hypothesis that has gained wide acceptance because it has survived rigorous investigation of its predictions. . . .''\cite{science_NAS}. 

In physics in particular, 
there is an extraordinary accordance between the results from calculations and results from measurements. 
This accordance between experiments and theory is very commonly used to validate the theory from where the prediction came, and sometimes is also used to argue in favor of a particular interpretation for such theory.
In the complex interplay between experiments and theory development, 
measurements have a central importance 
\cite{kuhn1961function}. 

Therefore, in the main trend of science, in one hand measurements have a tremendous weight in assessing the "trueness" of theories, but at the same time they are hardly analyzed, not a particular measurement but the concept and practice itself, just being taken for granted as a trusted way to obtain data.
This situation has been clearly stated in ~\cite{mari2005problem}, where it has received the name of "paradox of foundation" and is defined in these lines:
\begin{quote}
As all empirical sciences were asking measurement to play the foundational role of "protocol of truth" and Measurement Science accepted this function of delegate to deal with "pure data", measurement itself was forced to the paradoxical position of being at the same time the most empirically objective operation, because of its institutional tasks, and the most metaphysically based one, because of its conceptual foundation on the hypothesis of the existence of true values. We will call this clashing situation the "Paradox of Foundation" (PoF) for measurement.
\end{quote}

The main step in order to analyze measurements under a critical eye was taken at the end of nineteen century by Helmholtz~\cite{von1930counting}. 
He was followed by Campbell, Holder, Ellis and others~\cite{diez1997hundredI,diez1997hundredII,Ellis1968Basic,darrigol2003number}. 
These works are now considered part of the representational view of measurement\cite{krantz2006additive,suppes2007foundations,luce2007foundations}. 
In this view, at the heart of every measurement is a correspondence (homomorphism) between empirical actions and mathematical operations.

At the beginning of this century there have been proposals to go beyond the representational description~\cite{mari2000beyond,finkelstein2009widely}. 
This work contributes in this direction analyzing an important element of measurements: the measurement procedures. 
According to the International vocabulary of metrology~\cite{joint2008international}, a measurement procedure is a "detailed description of a measurement...". 
We propose a formal characterization of this key element in measurements using algorithms and its theoretical framework. 
This proposal emerges from the parallelism between the intersubjetive properties of both entities. 

Using an analogous argument to the halting problem in Turing machines, we draw limitations in the possible measurement and verification procedures.

\section{Measurements as Processes}

When we take a look to modern research laboratories around the globe, there is a very interesting feature that attracts the attention of people not related to the work in the lab: as years pass there is less human labor and more automated labor in the execution of experiments. 
Sometimes a whole experiment is started by pressing a button, then there is no other human intervention. 
This is a good feature to experimenters since leaves them more time to improve their experiments or design new ones.

In those completely automated experiments, computers play a fundamental role, inside them all the instructions needed to perform the experiment have been written previously. 
Such computers communicate with devices that allows them to "impose" conditions in the experiment, and also to verify them or "measure" a quantity, for instance some voltages or currents. 
Therefore, computers run the "logical part" or "instructions", giving orders to the different interfaces in order to perform such actions.

Since we live in times where technology is more and more part of our daily life, this situation can be seen as a mere consequence of such era and its huge  technological capabilities. 
We invite the reader to follow another interpretation of this situation. 

According to the International Vocabulary of Measurement (VIM) \cite{joint2008international} a \textit{measurement} is: 
\begin{quote}
  [A] process of experimentally obtaining one or more quantity values that can reasonably be attributed to a quantity.
\end{quote}
There are three important elements in this definition: measurement is a process, the end of which is to obtain a quantity value that is attributed to a quantity. Though these three elements are of importance for our analysis, we will focus on the first one: measurement is a {\it process}. 

\textit{Process} is a term used to name some actions in a specific order. For instance in dictionary {\it process} is defined as: "A series of actions or steps taken in order to achieve a particular end." \cite{stevenson2010oxford}. 
Examples of this can be found in daily life, a cooking recipe is a very common example of a process, is composed of {\it actions} like "set the oven to 150 degrees", "mix three eggs with sugar", and others. Putting together a prefabricated desk is also a process. 
These normally include a set of instructions that describe the process needed to assemble it.

Measurements, in the current conception of the scientific community, do fit as processes. 
They consist in a series of actions with a very specific end: to obtain the result of the measurement, the quantity value. 
They are composed of steps or actions. For instance, in order to measure the length of the table is necessary to use a ruler, put it parallel to the length of such table, one end of the ruler must touch one end of the table, then read the number, or closest number to the other end of the table. 
These steps can be considered the bricks that constitutes the measurement.  
It must be kept in mind, however, that following them only provides the quantity value.

These characteristics of measurement help us to understand the extensive use of computers in the modern lab. 
Since measurements are processes with a definite order, i.e. with some logical structure, it is possible to write that order in a computer which will execute the actions in the right sequence, when all the elements needed are provided.

In principle, all the actions required to perform a measurement can be described to other persons, talking to them, writing to them, etc. This description is called the \textit{measurement procedure}. According to the VIM\cite{joint2008international} a measurement procedure is: 
\begin{quote}
  [A] detailed description of a measurement according to one or more measurement principles and to a given measurement method, based on a measurement model and including any calculation to obtain a measurement result.
\end{quote}
In the scheme we've been developing, each action is related or described by an instruction in the procedure, usually written as an imperative sentence.  
Within this description the order in which the actions should be followed is also given.
The main idea is that every person with access to the measurement procedures and capable of executing every action required in that procedure will be able to perform the measurement.

Since not every process is a measurement, the actions which constitute a measurement should meet some restrictions. 
We have already mentioned that a measurement should be a process that returns a quantity value, i.e. a number and a reference, or even a formal entity (although that would be outside the definition). 
Some authors \cite{mari2017quantities} have pointed out other basic characteristics that a measurement should meet. 
For instance, a measurement should be not an opinion or guess, even if it is expressed by numbers. 
This condition can be directly translated to the actions which constitutes the experiment and the its structure: we cannot use opinion or guess in the individual actions of the experiment or in the flow of it.

An example of this situation is the instruction "write, in a one to then scale, the possibilities of storm today". 
This cannot be used as an instruction since uses the opinion of the person who is performing the experiment. 
A little change can turn this instruction in a valid one, for example, "ask to a person to write, in a one to ten scale, the possibilities of storm today". 
Since the experimenter's opinion is not involved in the execution of the experiment this is a valid instruction. 
We can also introduce these opinions or random processes in the flow of the process in a valid way, the instruction "decide which blood group will be examined" is not valid while "throw a dice to decide which blood group will be examined" is. 
This type of instructions should be accepted as valid instructions, since they are present in the practice of the scientific community. 
One common example are statistical proofs, where it is important to exclude the possibility of bias in an experiment.

As, by construction, these processes are defined by {\it mechanical} rules and actions, where any decision must be mechanical, they share this feature with algorithms.

There are other characteristics that link measurement and algorithms. Algorithms are made of processes that lack creativity. 
If there is a measurement procedure of a well established experiment, one that is already accepted by the community,  it is expected to have precise instructions that allows any person to perform it. 
Instructions like "figure out a way to connect this device with that one" cannot be part of the measurement procedure, because they require creativity from the performer of the experiment.

We have exhibited a parallelism between measurements and algorithms: both return results, both have detailed descriptions, measurements have measurement procedures, algorithms have their code, or symbolic representation that describes them. 
There is, maybe one difference between these concepts depending on what we mean by algorithm. 
An algorithm normally has only one result, the correct result. 
Depending on the repeatability of the experiment, measurements can have different results no matter how well specified are the conditions in the measurement procedure.  
We will return to this difference later.

\section{Effectivity in Measurements }

In the previous section we have described the parallelism between measurements and algorithms without specifying what we mean by an algorithm.  In this section this relation is analyzed using a formal definition of algorithm. 

Algorithms, in the modern view, are considered {\it effective procedures} \cite{cleland2002effective}.  
Effective procedures are those whose process can be executed immediately, at will, without obstacles and will always have a result. 
Write "hello" in a piece of paper is a widely accepted effective procedure, given that you have something to write and something to write on. 
Other example can be the measurement of the length of a table as we discussed earlier. 
Both examples can be executed at will, in any time, assuming that all the necessary resources are available.

On the other side we have examples of processes that are not effective: "wait until your aunt comes and say hi to her" is not effective.
While we can start it at will, i.e. start to wait, we have no certainty if the second action can be executed.
Maybe our aunt will never appear or maybe we do not have an aunt.  Other example is "divide 1 over 9 in the decimal system". Such division has an infinite decimal expansion so we can start to divide but we know for sure we will never end.

We have mentioned above the lack of creativity as a feature of processes in the last section, since a creative processes may or may not have a result. 
It can be analyzed also as a requirement for effectivity.  
For instance, in the previously mentioned procedure "figure out a way to connect this device with that one" is not an effective procedure, because a person could start to think how to achieve such task but there is no certainty of its success.

As measurements are expected to always produce results, when properly performed, in the present context we assume that all measurement procedures are effective, {\em i.e.} that every measurement procedure will end with certainty with a result.

The hypothesis of effectivity in measurement procedures is not explicit in the definition of the International Vocabulary of Metrology, but it is implicitly assumed by experimentalists around the globe. 
While it may be obvious in some communities, in the following formal analysis is important to state it explicitly.

\section{Formalization of effectivity and measurements}

\subsection{Algorithms}

Algorithms have a formal representation in modern math. As a consequence of the importance that algorithms had for mathematicians in the beginnings of the past century, they created a formal framework to describe them. 
Though not the first one, the approach used by Alan Turing is conceptually  simple and nowadays very popular. The formal objects he created are called \textit{Turing Machines}.

A Turing machine is composed of basic instructions,
similar to the series of actions described above. 
That is, Turing machines are descriptions of procedures or procedures schemas~\cite{cleland2002effective,cleland2004concept}, in the same manner that measurement procedures are descriptions of measurements. 
This instructions are designed to be performed by a very specific machine, a kind of printer,  that consist in a mechanism and a paper tape made of squares, each of which can only contain one symbol at a time. 
On this tape the machine can perform only three operations: read a symbol from a specified set, called the \textit{alphabet} of that machine, print a symbol from the alphabet and move to the adjacent squares.

The formal representation of an instruction consists of four symbols. For instance, the printing instruction is formally represented by:
\begin{align}
	q_i S_j S_l q_m ,
\end{align}
meaning that when the machine is in the state $q_i$ 
(loosely speaking, the state is the list of symbols which describes the actions the machine can execute
after reading a symbol from the tape.)
and reads the symbol $S_j$, it will print the symbol $S_l$ and change its state to $q_m$.

The moving operations are represented by:
\begin{align}
	q_i S_j R \, q_m \\
	q_i S_j L \, q_m 
\end{align}
In this case the interpretation is: if the state of the machine is $q_i$ and it reads the symbol $S_j$ then it will move to the right(left) and pass to the state $q_m$.

There are other basic considerations needed in order to understand how this instructions are executed by the machine. The idea is that the machine has its whole set of instructions "inside". It starts by convention in the state $q_0$.  
It reads the first symbol, then applies the instruction that matches that state and symbol, if doesn't have any, it halts, meaning that the calculation has ended.
Since a Turing machine is just a set of these instructions and the two first symbols decide which instruction is used in given situation, in a given machine there cannot be two different instructions whose first two symbols are the same.

Please note that, contrary to the terminology, Turing machines are not machines or its composition, they describe the process that a machine will execute. 
Turing machines are sets of instructions that represent processes. 
Again we like to stress the analogy between these machines and measurement procedures. 
While Turing machines describe algorithms, measurement procedures describe measurements.

In the big picture, the way used to formalize effective processes consists in selecting a small set of very simple processes which are considered effective and well defined. 
They should be so simple it is impossible to deny this two properties. 
With them it is possible to generate a greater set of processes that consists in combinations of the basic ones and the way they are combined is such that the property of effectivity is conserved. 
As a consequence the \textit{basic} procedures have a central role in the argument.  
Turing machines, with this restrictive set of instructions and rules, describe very well calculation procedures performed by humans.
However, as Cleland has pointed out~\cite{cleland2004concept}, the actions corresponding to this basic instructions (read, write and move), 
have no special place or properties. 
They are no different than the instructions found in a cooking recipe. These actions are different from others because the interested community
accept them as well-defined and effective.
The reason could be that they are familiar to all members of the community, or because, as Cleland said, it is easy in such community to train a person to do such actions in an unequivocal manner. 
Once the community accept them as basic effective procedures, their combinations will result in more effective procedures. 

\subsection{Experimental processes}

Having clarified this point, we consider a wider set of actions, powerful enough to represent other processes. 
They involve more complex actions than writing, reading and moving on a tape.
Of course, as we have discussed earlier, these set of actions have to be considered by the community of that field as effective and well-defined in the context their discipline. 
We call such set of actions the \textit{experimental possibilities} and are analogous to the printing, reading and moving of a Turing machine.

For instance, to impose a voltage between two cables can be see as a trivial task in the state of art of modern labs, and therefore can be considered as an action part of our experimental possibilities. 
However, we need to have in mind the huge number of hypothesis behind this statement: the cables are made of metal, are isolated, they are of reasonable size, the voltage will be on average the given value but will be some noise, etcetera. 
Such action may be considered trivial in the electronics community, where they are part of their every day practice, but not necessarily in the medicine community, or even the transmission lines community.

It is also true that sometimes, in given a community, it is possible to reduce an action to simpler ones, explaining the action or using an apparatus that simplifies the action. 
From the point of view of scientists in the 19th century, to impose a voltage of 10 volts in a pair of cables was not a trivial task at all, even with all the previous mentioned hypothesis.  
To impose a voltage is a trivial task for modern laboratories because they have voltage supplies with which it is possible to translate such action into simpler ones, like pressing buttons and connecting wires. 
Removing such supplies can transform an effective procedure into a non-effective one, if there is no other way to impose such voltage with the other equipment in the lab or there is no knowledge of how to do it since, as we've mentioned above, creative processes are ruled out. 
Cleland has also discussed this issue~\cite{cleland2004concept}:
\begin{quote}
The success of human beings in following imprecisely described instructions is a product of training coupled with a shared repertoire of basic bodily actions such as moving a finger or rotating a wrist.
\end{quote}
 i.e.  all actions can be reduced to some set of actions she called "bodily actions".
For our present purposes it is enough to consider the set of basic experimental possibilities, those accepted by the community in which the procedures are executed.

\subsection{Experimental procedures}

With this considerations in mind, and specially due to the nature of Turing machines as "procedure schemas", as Cleland puts it, but using a more complex set of basic actions, it is possible to use such machines in order to represent measurement procedures. 
We will formally represent a measurement procedure as a set of instructions, just like a Turing machine. 
This instructions, however will use a wider set of actions. 
For instance, an instruction can be represented in the following way:
\begin{align}
q_i S_j I_k q_l q_m
\end{align}
meaning that: "if you are in the state $q_i$ and you read a symbol $S_j$ then perform the action $I_k$ and pass to the state $q_l$. 
In case the action $I_k$ cannot be performed, move to the state $q_m$". 
This description only leaves the actions $I_k$ as abstract actions that should fulfill the conditions discussed above.

Many actions need to receive information, to "see" something other than the tape.
We call them \textit{reading actions} $V_k$, they are actions that verify some binary statement. For instance, "the second digit is seven". Depending if this statement is true or not it is possible to make decisions. 
They are represented in this type of instruction:
\begin{align}
	q_{i} S_{j} V_{k} q_{l} q_{m} q_{p},
\end{align}
whose meaning is: "if you are in the state $q_i$ and you read a symbol $S_j$, then perform the reading action $V_k$ and pass to the state $q_l$, if 
the statement is true, and $q_m$ if the statement is untrue. 
In case the action $V_k$ cannot be performed, move to the state $q_p$".

Here we have kept the tape of the Turing machines, as a representation of a "piece of paper" that will allow to write results. 
They can be used in later steps to perform calculations, since the mere definition of measurement procedure requires such. 
Therefore these machines can perform actions in order to execute the measurement, but restrict themselves to write, read and move when only formal calculation are required.
\footnote{
  This calculations, by the way, can be performed without any knowledge of the math beneath them since can be always reduced to the mechanical operations of the machine. 
}

The presence of the tape allows a formal definition the measurement result, as the string of characters leaved in the tape once the machine stopped. Note that the measurement result is by definition a \textit{quantity value} that is composed, again by definition in the vocabulary, by a number and a reference, e.g. 10km. 
In the present context it is a mere sequence of symbols, i.e. "10km", without the meaning of number or reference. 
In this way the algorithmic description includes generalized measurement process, where the result can be a nominal (qualitative) quantity value ~\cite{joint2008international}, like the sex of an animal or the colour of a spot test in chemistry.
It is important to stress that this description only applies to the measurement procedure. 
The association between quantity and quantity value is outside its scope.

Having specified the result of a measurement, we can discuss its repeatability.
A machine is repeatable if every time the procedure is executed it gives the same result, the same string of symbols. 
In those cases in which the result is a number and a reference, uncertainties and noise must be considered. 
Adding to the instructions in the procedure to only report a definite number of significant figures, a reasonable set of repeatable measurement procedures can be obtained. 
This is a very narrow, algorithmic definition of repeatablility,

We end this section with a summary of the formal concepts we have introduced. 
To represent measurement procedures, we use formal objects called Turing machines. 
These objects have a set of symbols and a endless tape. 
Using them they can perform operations: write, read and move in the tape.
In order extend their possibilities we add basic processes called actions, that extend the possible instructions for the machine.

While a regular Turing machine consists of three types of instructions \footnote{We can add the oracle as a fourth instruction~\cite{davis1982computability}.}:
\begin{align}
	q_i S_j S_k q_l,\\
	q_i S_j R q_l,\\
	q_i S_j L q_l,
\end{align}
the proposed formal model has two more, that is five possibilities:
\begin{align}
	q_i S_j S_k q_l,\\
	q_i S_j R q_l,\\
	q_i S_j L q_l,\\
	q_i S_j q_l q_m,\\
  q_i S_j I_k q_l q_m,\\
  q_i S_j V_k q_l q_m q_n,
\end{align}
where $I_k$ represents the $k-th$ possible action of the machine, and $V_k$ reading actions. These are well-defined and effective actions accepted by the community in which these procedures are designed and executed. 
The set of all actions performed by a machine $\{I_k,V_j\}$ are called its experimental possibilities. Finally, if the result of executing a machine $M$ with an input $e$, is the same every time, the machine is called repeatable and the result is denoted $M(e)$.

The association of modern computers with Turing machines allows for a generalization of their actions, to include the manipulation of punched cards, magnetic tapes, electronic devices, etc. They are referred as the {\em hardware}, whose detailed description is not required if it is guaranteed that they can realize the required actions. 
In a measurement procedure, the symbols describe the manipulation of a variety of systems whose properties are measured, and of the devices employed in the procedure, whose outputs are read. The automatization of many measurements helps to visualize these actions as simple extensions of Turing machines.

\section{Consequences of algorithmic formulation of procedures.}

Up to here we have discussed the relation between measurement procedures and algorithms, describing the later as Turing machines. 
Employing them we formally defined the result of a measurement and its repeatability. Now we are interested in the formal properties that can be proved using such definition.

The most famous and important result in relation with Turing machines is the halting problem. 
It consists in determining if a machine with a given input will ever halt. 
This is important to our discussion since halting implies having a result and we have stated that a measurement must be effective, i.e. must end with certainty with a result.

We pose the question: Is it possible to determine whether a given procedure corresponds to a valid measurement process, i.e. it {\em always} has a result? 
This effective criteria, if it exists, should be another process whose input is a procedure, and the output is yes, it is a measurement process, or no, it is not.  
We show here that it is not possible to have such criteria.

\subsection{The Halting Problem}

This is a brief and informal description of the Halting Problem~\cite{kleene2002mathematical, davis1982computability}.

\begin{enumerate}
  \item We start assuming that there is a {\em decisive} process 
  that allows  to decide if a given procedure halts, i.e. gives a result. 
  Such process takes a procedure, i.e. a formal description of the instructions, as its input, applies some actions and returns 'H' if the procedure has a result (halts) and 'N' in the other case.

  \item A second process is built employing the first. It works in this way: 
  \begin{itemize}
    \item It has both the target procedure and the decisive algorithm as its inputs.
    \item It reads the output of the decisive algorithm applied to the procedure.
    \item If the output is 'N' it gives a result, which can be any, and halts.
    \item If the output is 'H' it never gives a result, continues to perform actions idly.
  \end{itemize}

  As a result, this second, combined process halts if and only if its target procedure doesn't.

  \item 
  Since the combined process has a procedure, i.e. a description, just like any other, it can be employed as the target procedure. 
  In this case "The combined process halts if and only if the combined process doesn't". 
  This is a clear logical contradiction.  
\end{enumerate}

The hypothesis that lead to the contradiction was that there is a process that can always decide if a given procedure halts. 
It follows that such process cannot exist. 
No process can always decide if a given procedure halts.

\subsection{Valid measurements}

\begin{figure}
	\centering
	\includegraphics[width=0.2\textwidth]{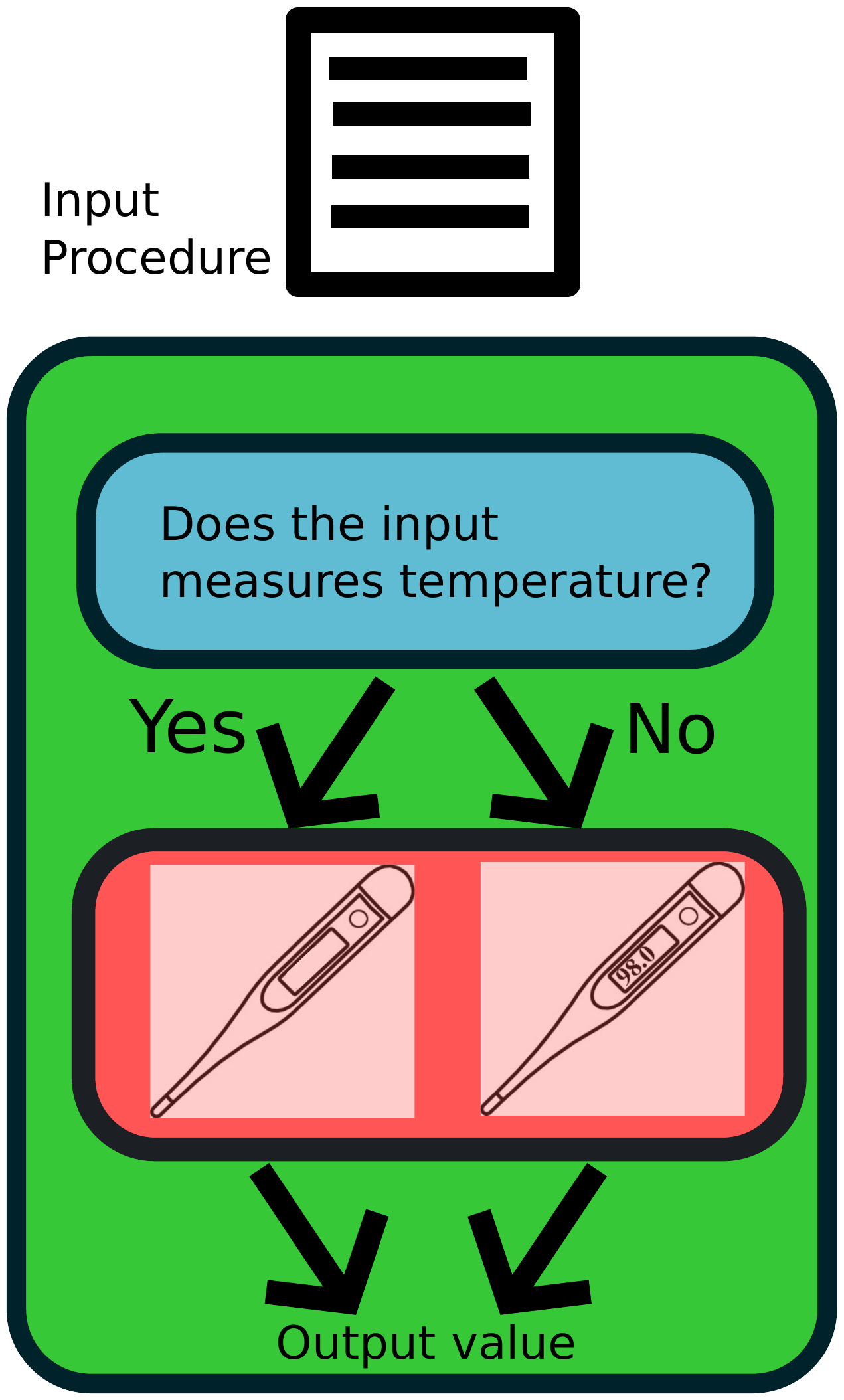}
	\caption{The assumption of the existence of the blue experimental procedure leads to a contradiction.}
	\label{prueba}
\end{figure}

In this section, we extend the above result to prove that it is {\bf not} possible to determine if a given procedure measures a given quantity. 
To simplify the discussion, let's assume that there is a tentative procedure to measure temperature and want to determine if it is a valid measurement procedure. 
The condition is that it must be effective, give a result in each occasion it is employed and coincide with  the reference measurement procedure \cite{joint2008international}.
The demonstration follows the same steps as above. The whole process if represented in  Fig. \ref{prueba}.

\begin{enumerate}
  \item We start assuming that there is a {\em verification process} 
  that allows to decide if the tentative procedure measures temperature. 
  It takes the tentative procedure as its input and returns 'Yes' if the procedure measures temperature and 'No' in the other case. It is represented as the {\em blue} process in  Fig. \ref{prueba}.

  \item A second, {\em green} process is built employing the first. It works in this way: 
  \begin{itemize}
    \item It has the tentative procedure as input and uses the verification process.
    \item It reads the output of the verification process applied to the procedure.
    \item If the output is 'No' it measures the temperature employing the reference procedure, prints the result and halts.
    \item If the output is 'Yes' it turns off the tentative procedure, prints 'no temperature has been measured' and halts.
  \end{itemize}

  The last two steps are represented as the {\em red} process in  Fig. \ref{prueba}.
  As a result, this green process measures the temperature if and only if its target procedure doesn't.

  \item 
  The green procedure is fed with its own procedure as input. Then,  
  "The green process measures the temperature if and only if the green process doesn't". 
  Again we arrive to a  logical contradiction.  
\end{enumerate}

The contradiction proves that the first hypothesis was false: there cannot exist a {\em verification process} that allows to {\em always} decide if the tentative procedure measures temperature. In general, it is {\bf not} possible to determine if a given procedure corresponds to a valid measurement process, i.e. it {\em always} has a result. 

\subsection{Formal demonstration}

The argument we have just draw is a direct application of the diagonal method. In a way, is a carbon copy of Turing's {\it halting problem}, Rice theorem or other related results~\cite{kleene2002mathematical, davis1982computability}. 

Lets define the set of process $\Omega$ as all the effective processes we've discussed earlier. Mathematically it is the set of machines that halt, so they have a result with certainty, though such result does not have to be repeatable. In other words, a process $M\in \Omega$  will halt in every execution, but not necessarily return the same result. 

Above we insisted in the relation between a process and its procedure, formally a process $M$ has a description or procedure that we will denote as $\quine M$. Please note that $\quine M$ is just a string of symbols that allows any trained person (or a computer with an appropriate code)
to execute the process $M$.

Suppose that there is a repeatable process $blue\in \Omega$ capable to decide if another procedure measures temperature when fed with an input $e$. That is $blue(\quine M, e)$ returns '1' if $M(e)$ always measures temperature and '0' in the other case.
$blue$ should be a repeatable
\footnote{This is necessary because there are non-repeatable procedures that sometimes measure temperature and sometimes other property but the election is done in a random fashion.}
process if we want the right answer with certainty. $M(e)$, though not necessarily repeatable, should measure temperature always in order to satisfy blue. 
Finally, we use $T(M(e))$ to denote that $M(e)$ measures temperature, i.e. $T(M(e))$ is either true or false. 

Using the above elements, we can write the property of the $blue$ process as:
\begin{align*}
&if M \in \Omega \implies \\
&blue(\quine M,e)='1' \iff T(M(e)).
\end{align*}

For the next step in our argument we need a $green$ process with the particular property  "measures temperature only when its input doesn't".  We accomplish this using the procedures $blue$ and $red$. This can be represented by:
\begin{align*}
&if M \in \Omega \implies \\
&T(green(\quine M,e)) \iff \neg T(M(e)).
\end{align*}

Next, we feed the process with its own procedure by defining the process $G(\quine M)=green(\quine M,\quine M)$. Therefore the last formal statement can be rewritten as:
\begin{align*}
&if M \in \Omega \implies\\
& T(G(\quine M)) \iff \neg T(M(\quine M)).
\end{align*}

The last step in order to get the desired contradiction is to take the special case $M=G$:
\begin{align*}
&if G \in \Omega \implies \\
&T(G(\quine G)) \iff \neg T(G(\quine G)).
\end{align*}

The conclusion is that the procedure $G$ cannot exist or it cannot be in $\Omega$. 

\subsection{The hypothesis involved}

We have assumed the existence of the processes $blue$ and $red$.  
But only one of these machines can exist, the existence of both lead us to a contradiction.
For the $red$ procedure to exist, there must be an alternative procedure available to be employed when the output of $blue$ is '1'. 
For this reason, this demonstration is valid when a standard procedure has already been accepted by a community to measure a given quantity. 
It does not apply when new quantities are introduced. 
It must also be possible to 'turn off' the input procedure. 
This is very simple, just to instruct $red$ to do not use it in this case is enough. 
It is the $blue$ procedure the one which cannot exist.

There is another important aspect of the proof that may awake some objections, we have only considered procedures that have an argument. 
There are lots of procedures that have an argument, using it they decide which actions should be performed. 
But, there are numerous measurement procedures that does not need an argument in order to perform its task. 
Are our conclusions valid for these procedures?
%Are this procedures out of our argument? It turns out that there is scape from our argument. 
Consider a process $M$ with a particular input $e$. 
%Now there is 
It is always possible to design
another process that  doesn't take any arguments. It first prints the particular input $e$ and then, immediately, performs all the actions $M$ would perform normally. 
Therefore, for every process with a particular input, there is another process without input that reproduces exactly the process followed by $M$ with input $e$. 
If we had a measurement procedure capable of solving whether or not a process without arguments measures temperature, we will be capable of solving the problem for all the processes with a particular input. 
Since we've already discarded that option, the processes without arguments cannot have a effective procedure to decide if them measure temperature or not.

It is actually possible to extend this method, the diagonal method, to other properties of interest in the realm of measurement processes and procedures. 
There may be other important verification 
processes that also had some limitations due to the nature of the property they verify, in the same manner Rice theorem applies for a big set of properties \cite{hopcroft2008introduction}. 
This will be subject of other work.

\section{Conclusions}

Along this article, we have shown that measurement procedures are processes, which can be described as algorithms. 
As every measurement is expected to produce a result, we concluded that these processes must be effective. 
It allowed us to describe measurement procedures as generalized Turing machines. They have results, chains of symbols which are generalized quantity values.  

We introduced the experimental possibilities, the set of actions considered by the community of that field as effective and well-defined in the context their discipline. 
They are analogous to the printing, reading and moving of a Turing machine.  
Following the same logical construction of the {\em halting problem}, we proved that 
it cannot exist an effective algorithmic process to determine
if a tentative procedure describes a valid measurement process. 
In general, this relevant question must be agreed
 by the community involved.  

\section*{Acknowledgements}
We acknowledge funding from DGAPA- UNAM project IN109417.

\end{document}